\documentclass[reprint,amsmath,amssymb,aps]{revtex4-1}

\usepackage{graphicx}
\usepackage{dcolumn}
\usepackage{bm}

\begin{document}


\title{Atlas for the properties of elemental 2D metals}

\author{Janne Nevalaita}
\author{Pekka Koskinen}%
 \email{pekka.koskinen@iki.fi}
\affiliation{%
Department of Physics, NanoScience Center, University of Jyv{\"a}skyl{\"a}, 40014 Jyv{\"a}skyl{\"a}, Finland
}%

\date{\today}

\begin{abstract}

Common two-dimensional (2D) materials have a layered 3D structure with covalently bonded, atomically thin layers held together by weak van der Waals forces. However, in a recent transmission electron microscopy experiment, atomically thin 2D patches of iron were discovered inside a graphene nanopore. Motivated by this discovery, we perform a systematic density-functional study on atomically thin elemental 2D metal films, using 45 metals in three lattice structures. Cohesive energies, equilibrium distances, and bulk moduli in 2D are found to be linearly correlated to the corresponding 3D bulk properties, enabling the quick estimation of these values for a given 2D metal and lattice structure. In-plane elastic constants show that most 2D metals are stable in hexagonal and honeycomb, but unstable in square 2D structures. Many 2D metals are surprisingly stable against bending.

\end{abstract}

\maketitle


\section{Introduction}

Covalently bonded 2D materials have a wide range of exceptional properties~\cite{novoselov04, novoselov05, geim07,geim09}. Graphene, for example, is extremely strong~\cite{lee08}, has high thermal conductivity~\cite{balandin08} and high charge carrier mobility~\cite{bolotin08}, and can exhibit the Quantum Hall Effect~\cite{zhang05,novoselov07}. Many other 2D materials, such as hexagonal  boron  nitride and transition  metal chalcogenides, are also investigated for their exceptional properties and promising applications~\cite{butler13, miro14, bhimanapati15, zou15, lin16}. The structure of these 2D materials is related to the nature of their bonding. In the three-dimensional (3D) bulk, the covalently bonded 2D materials consist of tightly bound layers that are connected by weak van der Waals forces and that can be exfoliated even into single free-standing layers~\cite{coleman11,nicolosi13}. However, 2D materials with metallic boding are largely unexplored. Some metals are known to form 2D structures on supports, including K on graphene~\cite{yin09,yin15a,yin15b}, Pb and In on Si(111)~\cite{zhang10}, Hf on Ir(111)~\cite{li13}, Sn on Bi$_2$Te$_3$(111)~\cite{zhu15}, Rh on polyvinylpyrrolidone~\cite{duan16}, and Ga on multiple substrates~\cite{kochat17}. Compared to covalent 2D materials, metallic bonding prefers close-packed structures, not layered ones. Therefore, free-standing 2D materials with metallic bonding have remained elusive. 

Still, 2D materials with metallic bonding have numerous potential applications~\cite{ling15}, including catalysis~\cite{deng16} and gas sensing~\cite{pan14}, which makes them an inviting research subject. Also recent literature encourages studying 2D metals further. For example, in a recent experiment an atomically thin iron membrane was grown inside a graphene nanopore~\cite{zhao14}. The membrane appeared to be iron atoms in a 2D square lattice structure. Later theoretical works suggested that the iron patches are more stable as carbides, containing also carbon~\cite{shao15,chen17}. Similarly, a free-standing monolayer thick zinc oxide~\cite{quang15} and copper oxide~\cite{yin17} membranes have been observed inside a graphene nanopores. In addition to these 2D structures, theoretical works have predicted the existence of stable 2D gold~\cite{yang15b}, silver~\cite{yang15}, and copper~\cite{yang16} membranes. Further, a computational study found that Au membranes may form in graphene nanopores by way of only small energy barriers for Au diffusion~\cite{antikainen17}. Previous works also show that Au~\cite{amft11} and Pt~\cite{kong06} readily diffuse on graphene. While these examples are about solid 2D structures, simulations have predicted even the existence of a 2D liquid~\cite{koskinen15}. A gold membrane in a graphene nanopore was investigated with molecular dynamics simulations and a solid to liquid phase transition was observed. Later theoretical calculations extended the prediction of 2D liquid phase also to Cu, Ag, and Pt~\cite{yang16b}. This phase transition is facilitated by the flexible metallic bonding, which increases the mobility of the atoms compared to materials with directional and rigid covalent bonds. Therefore, materials with flexible metallic bonding have potential to establish a new class of 2D materials with novel properties. Alas, only a few elemental 2D metals have been studied; a systematic investigation with many elements is missing.

\begin{figure}
\includegraphics[width=8.6 cm]{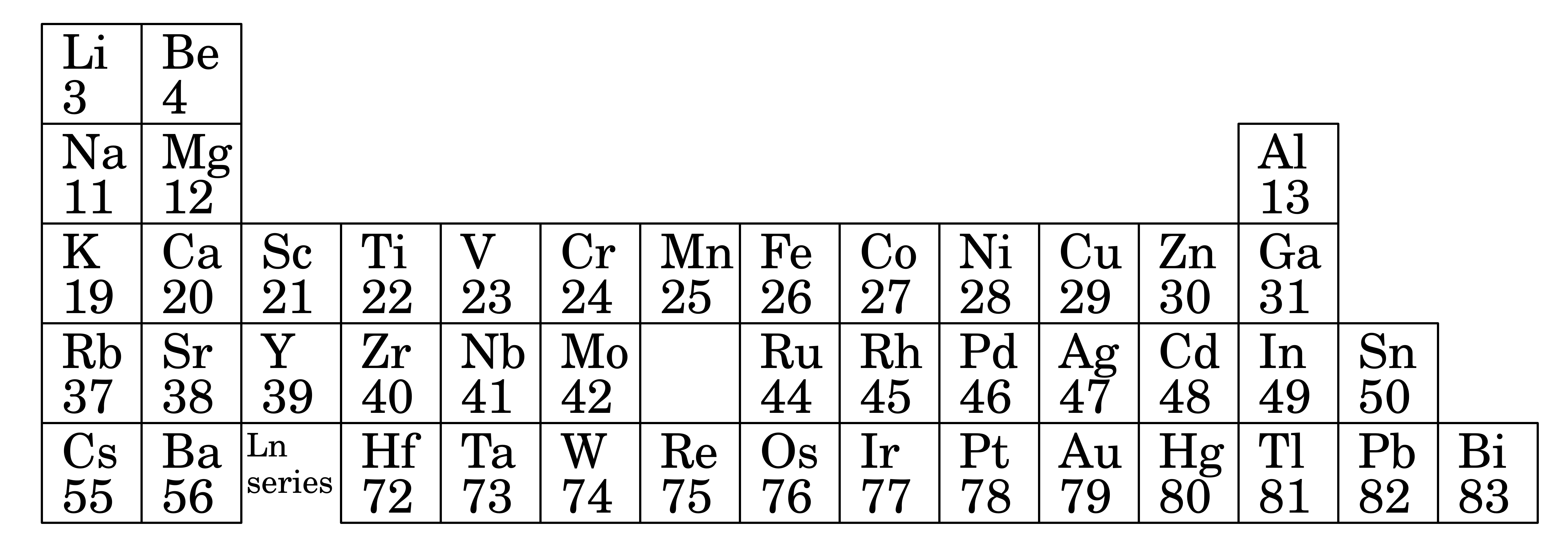}
\caption{Part of the periodic table with the studied metals. Chemical symbol and proton number are given for each element.}
\label{fig:metals}
\end{figure}

In this paper, we provide this missing systematic investigation. We present a  density-functional study of 45 elemental 2D metals (\mbox{Fig. \ref{fig:metals}}). Each element is considered in hexagonal, square, and honeycomb structures. These structures are chosen to obtain atoms with different coordination numbers. We compare the cohesive energies, bond lengths, and elastic parameters between the 2D structures. We find that many of the properties of the 2D metals are inherited from the corresponding properties of 3D bulk. Therefore, by using the well-known 3D bulk properties, the cohesive energies, bond distances, and bulk moduli can be quickly estimated for a given 2D structure and metal. The same properties correlate linearly also among the studied 2D lattice structures.

\section{Computational methods}

\begin{figure}
\includegraphics[width=8.6 cm]{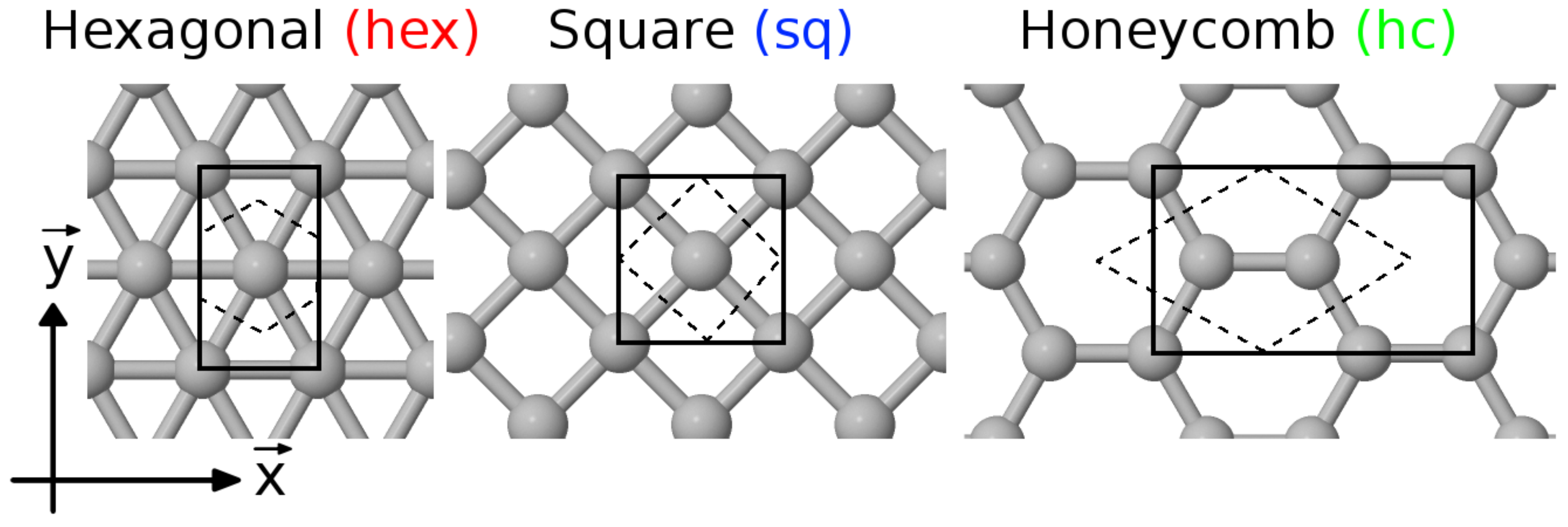}
\caption{Schematic representations of the 2D lattices. Solid line indicates the computational unit cells and dashed lines the Wigner-Seitz cells.}
\label{fig:lattice_sketch}
\end{figure}

The calculations, based on density-functional theory (DFT), were done in the plane-wave mode of the GPAW-code~\cite{mortensen05,enkovaara10}, using an \mbox{800 eV} plane-wave cut-off energy and default setups. Since the calculated systems were hypothetical, a non-empirical exchange and correlation functional was preferred. The Perdew-Burke-Ernzerhof (PBE) exchange and correlation functional~\cite{perdew96} is non-empirical, computationally inexpensive, and reproduces the bulk cohesive energies well~\cite{schimka13}. Although the PBE functional is known to reproduce bond lengths less accurately, we expect that the general trends are not that sensitive to possible errors in individual systems. Therefore, the PBE functional was used for all calculations. Three different 2D structures were studied, the hexagonal (hex), square (sq), and honeycomb (hc) lattices (\mbox{Fig. \ref{fig:lattice_sketch}}). The hexagonal and square lattices were modeled using two atoms and the honeycomb lattice using four atoms in the unit cell. All calculations had \mbox{5 \AA} vacuum regions for non-periodic directions. The flat 2D structures used a 12$\times$12$\times$1 and the 3D bulk used a 12$\times$12$\times$12 Monkhorst-Pack k-point sampling~\cite{monkhorst76,pack77}. Bent structures used a 1$\times$12$\times$1 sampling. All calculations were spin-polarized and convergence was checked with respect to vacuum layers, $k$-point grids, and plane-wave cutoff.

\section{Results}

\subsection{2D cohesive energies correlate between structures}

We begin by considering the cohesive energies of the 3D bulk and the three ideal 2D structures for all 45 metals. Most structures are nonmagnetic so we focus on other properties, beginning with cohesive energy. The cohesion is determined by calculating the energies of systems with ideal bonding angles and minimum-energy bond lengths. These are then subtracted from the energies of free atoms. The resulting values of 2D structures are compared relative to one another and to the 3D cohesion. The cohesive energy is defined as

\begin{equation}
E_{coh} = E_{atom} - E/N,
\end{equation}
where $E_{atom}$ is the energy of a free atom and $E/N$ is the energy of the structure divided by the number of atoms in the unit cell. With this definition, the larger the cohesion energy, the more energy is required to sublimate the system. In order to validate the computational method, the bulk cohesive energies are calculated for all 45 metals and compared to the corresponding experimental values~\cite{kittel}. As expected, the calculated 3D bulk cohesive energies follow the experimental values well, with a mean absolute error of \mbox{0.28 eV}. This indicates that the used numerical method is sufficiently accurate to identify trends between 2D geometries. The mean absolute error for 3D bulk cohesion is also in line with a previously reported value of \mbox{0.24 eV~\cite{schimka13}}.

\begin{figure}
\includegraphics[width=8.6 cm]{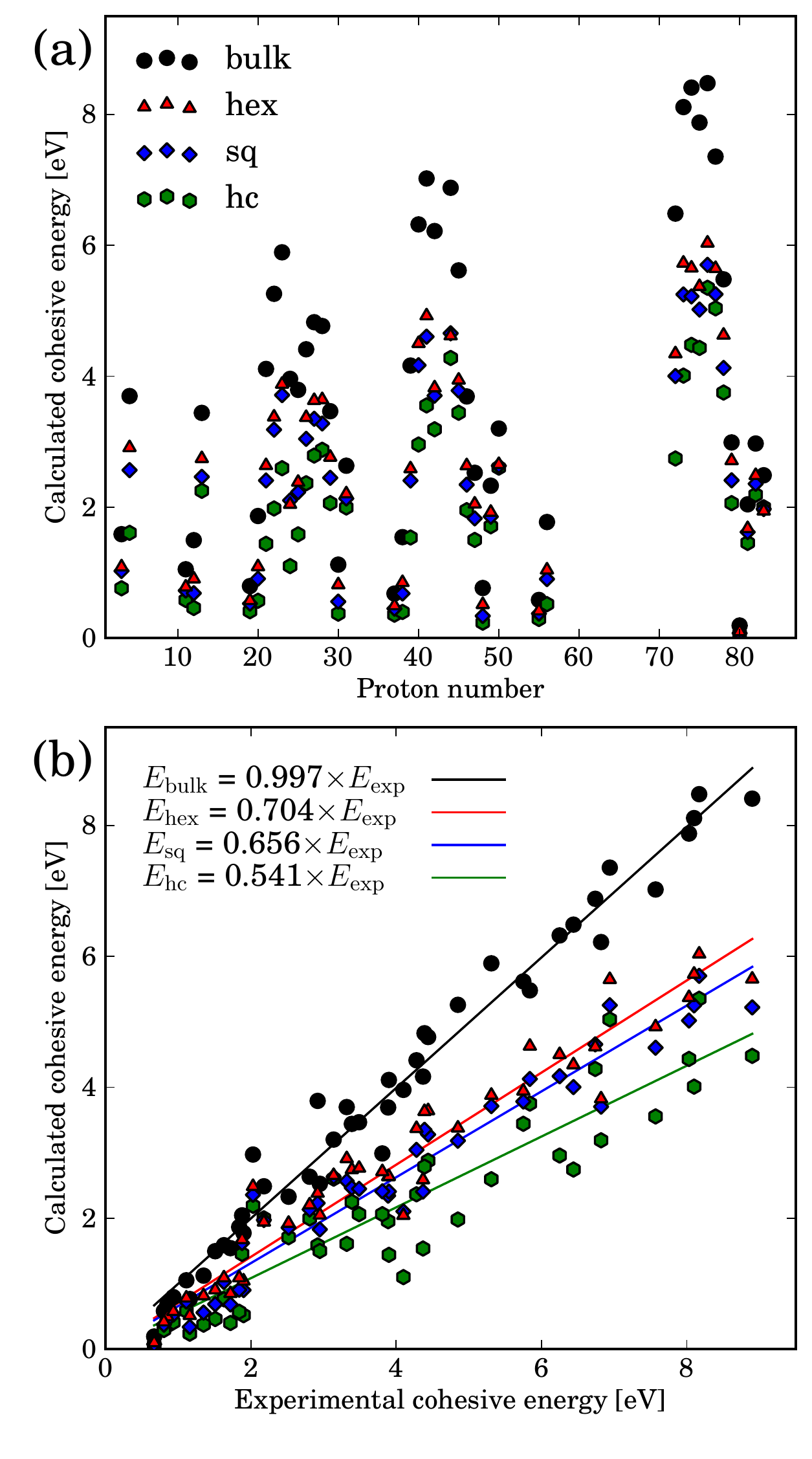}
\caption{Calculated cohesive energies of 3D bulk, and hexagonal (hex), square (sq), and honeycomb (hc) 2D lattices as a function of proton number (a) and experimental 3D bulk cohesion (b). Lines show linear fits between experimental and calculated values. Experimental cohesive energies are from ref.~\onlinecite{kittel}.}
\label{fig:E}
\end{figure}

The calculated cohesive energies for all considered structures and metals are shown in \mbox{Fig. \ref{fig:E}}. They range from nearly zero (Hg) to almost \mbox{9 eV} (5d metals W and Os). The weak cohesion of Hg is in part caused by its notoriously strong electron correlations, difficult to capture by any functional~\cite{gaston06}. Nevertheless, as discussed before, the PBE functional is otherwise expected to capture the correct trends in cohesion energies. The cohesion is, in general, the highest in the middle of the transition metal series. This can be rationalized by a simple model~\cite{ziman69}: Assuming that the energy of the metal can be approximated as a sum of single particle energies, the cohesive energy can be written as

\begin{equation}
E_{coh} \approx 2 \int^{\epsilon_f}_{-\infty}\mathrm{d}\epsilon(\epsilon_0-\epsilon)\rho(\epsilon),
\label{eq:Ecoh}
\end{equation}
where $\epsilon_f$ is the Fermi energy, $\epsilon_0$ the energy of the d-state for the free atom, and $\rho$ the density of d-states. If one further assumes that the density of d-states is constant $\rho_0$ in an energy range $w$ located symmetrically around $\epsilon_0$, equation~\eqref{eq:Ecoh} gives

\begin{equation}
 E_{coh} \approx -\rho_0(\epsilon_f^2-w^2/4),
\label{eq:Ecoh_approx}
\end{equation}
with zero energy chosen so that $\epsilon_0 = 0$. The number of d-electrons $N_d$ is obtained by integrating the density of d-states to the Fermi energy

\begin{equation}
N_d = 2\int^{\epsilon_f}_{-w/2}\mathrm{d}\epsilon\rho_o=2\rho_0(\epsilon_f+w/2)
\label{eq:Nd}
\end{equation}
and the number of electrons in a full d-band by a similar integral over the entire width of the d-band

\begin{equation}
N_{\mathrm{full}} = 2\int^{w/2}_{-w/2}\mathrm{d}\epsilon\rho_o=2\rho_0w.
\label{eq:Nfull}
\end{equation}
By combining the equations~\eqref{eq:Ecoh_approx}, \eqref{eq:Nd}, and \eqref{eq:Nfull}, we obtain a parabolic relation for $E_{coh}$ as a function of $N_d$

\begin{equation}
E_{coh} \approx \frac{w}{2}\Big(N_d-\frac{N_d^2}{N_{\mathrm{full}}}\Big)
\end{equation}
with the maximum cohesion in the middle of the d-series. While this simple model can be used to rationalize the qualitative trend in the cohesive energies, it fails to produce quantitative agreement. For example, the model holds better for the 4d- and 5d-series than for the 3d-series. The deviation in the 3d-metals can be attributed to strong Coulomb correlations, which cause band splitting~\cite{kajzar77}. In general, the simple s-metals have lower cohesion than the d-metals due to the lack of d-electron contribution to bonding.

Let us next look more closely how the cohesion energy is affected by the 2D structure. The energies of the calculated 2D structures correlate linearly. The calculated cohesive energy of the square lattice $E_{sq}$ is an approximate function of the cohesive energy of the hexagonal lattice $E_{hex}$, with a function \mbox{$E_{sq} = \alpha\times E_{hex}$} and a fit parameter $\alpha$. Values for the fit parameters between the different 2D structures are as follows: \mbox{$E_{sq} = 0.932\times E_{hex}$}, \mbox{$E_{hc} = 0.773 \times E_{hex}$}, and \mbox{$E_{hc} = 0.831 \times E_{sq}$}. Most important, the 2D cohesive energies can be linearly correlated to the experimental 3D bulk cohesive energies (\mbox{Fig. \ref{fig:E}b}). While the emerging energy correlations are not perfect, they well indicate the general trends, which are as follows. First, the cohesion is the strongest for the 3D bulk structures, as expected.  Second, generally the hexagonal geometry is the most stable and the honeycomb lattice the least stable. The square lattice cohesions are somewhere in between hexagonal and honeycomb cohesions. Previous studies on 2D iron~\cite{thomsen15}, silver~\cite{yang15}, and gold~\cite{yang15b} membranes obtained the same relative ordering. A more recent study also found the close packed hexagonal Au lattice energetically more stable than the honeycomb lattice~\cite{liu17}. Third, despite the dramatically reduced coordinations, the most stable 2D structure has about \mbox{$70\ \%$} of the 3D bulk cohesion. After all, most of the metals have 12 nearest neighbors in 3D bulk, but only six nearest neighbors in the hexagonal 2D lattice. Similarly, the square lattice has about \mbox{$65\ \%$} of the bulk cohesion with four nearest neighbors and the honeycomb lattice about \mbox{$54\ \%$} with only three nearest neighbors. Therefore, the average bond in a 2D structure is stronger than in the 3D bulk.

\begin{figure}
\includegraphics[width=8.6 cm]{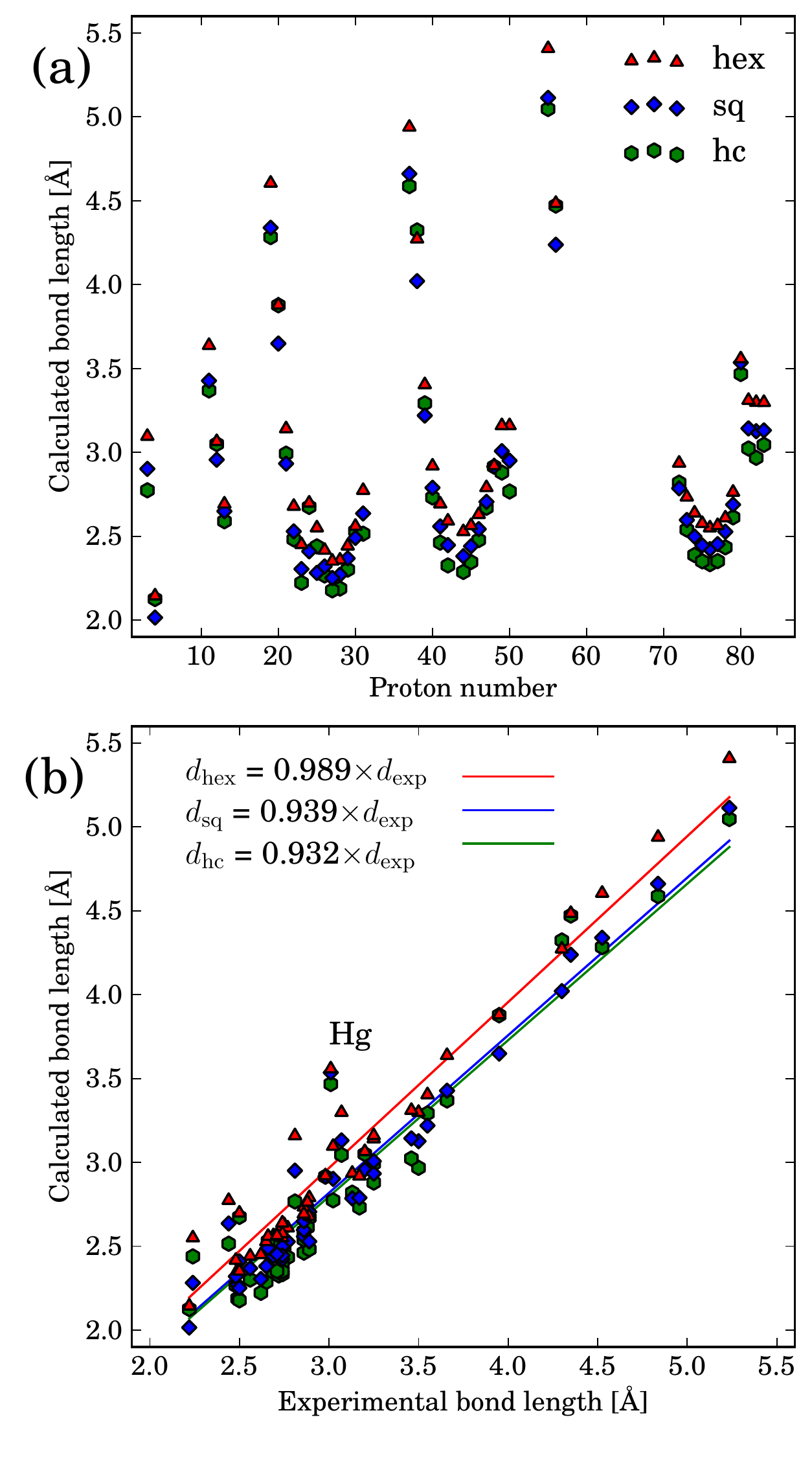}
\caption{Calculated bond lengths for hexagonal (hex), square (sq), and honeycomb (hc) 2D structures as a function of proton number (a) and experimental 3D bulk bond length (b). Lines show linear fits between calculated and experimental values~\cite{kittel}.}
\label{fig:d}
\end{figure}

\subsection{Equilibrium bond lengths correlate between structures}

Let us continue to study the bond lengths of the metals in the three ideal 2D structures. The reported bond lengths correspond to the lowest energies and range from $2$ to \mbox{$5.5$ \AA} (\mbox{Fig. \ref{fig:d}}). Comparison to the previous section shows that, while near the middle of the d-series the cohesive energies have maxima, the bond lengths have minima. This is in agreement with the maxim of stronger bonds being shorter~\cite{linus54}. Also, the changes in the interatomic distances have a parabolic shape for all d-series. Just as the cohesive energies, the bond lengths correlate linearly between the 2D structures. Performing fits as in the previous section, we obtain the values \mbox{$d_{sq} = 0.950 \times d_{hex}$}, \mbox{$d_{hc} = 0.943 \times d_{hex}$}, and \mbox{$d_{hc} = 0.992 \times d_{sq}$} for the bond lengths. For most metals, the bonds are the longest in the hexagonal lattice, intermediate in the square lattice, and shortest in the honeycomb lattice.

The bond lengths again correlate to the experimental 3D bulk values (\mbox{Fig. \ref{fig:d}b}). Hg deviates from the general trend, again perhaps due to inaccurate description of electronic correlation. As a trend, the bond lengths are the longest when the number of nearest neighbors is the largest. In 2D, the number of nearest neighbors is six for hexagonal, four for square, and three for honeycomb 2D lattices. In 3D, the number of nearest neighbors for most metals is 12. While the cohesion is reduced by about \mbox{30 \%} from bulk values, the bond lengths are only about \mbox{1 \%} shorter in the hexagonal 2D geometry. It seems that, while metals do not prefer layered structures, the geometries of the individual metal layers do not change much as the thickness of the metal reduces to a monolayer. For the square and honeycomb lattices, however, the bond lengths do shrink considerably compared to the 3D bulk. Yet the bond lengths in square lattice are only \mbox{1 \%} longer than in honeycomb lattice. This could reflect the smaller change in the number of nearest neighbors: When the structure is changed from the hexagonal lattice to the square lattice, the number of nearest neighbors is reduced by one third, but when the structure is changed from the square lattice to the honeycomb lattice, the number of nearest neighbors is reduced only by one fourth. In the previous section we observed that the 2D bonds are stronger than the 3D bonds. The shorter bond lengths for the 2D structures are in line with this observation.

\begin{figure}
\includegraphics[width=8.6 cm]{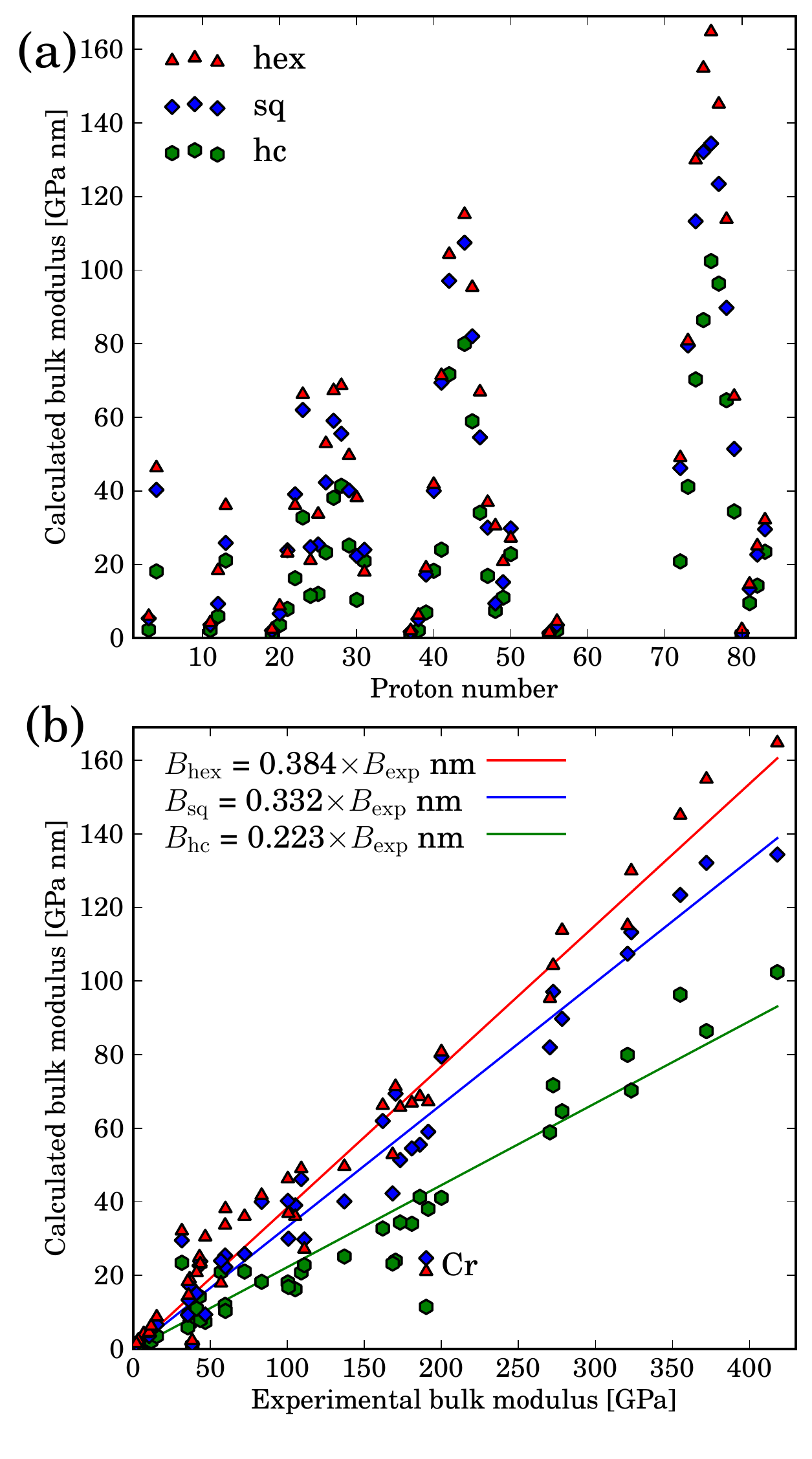}
\caption{Calculated 2D bulk moduli for hexagonal (hex), square (sq), and honeycomb (hc) structures, as a function of the proton number (a) and experimental bulk moduli (b). Lines display the linear fits between calculated and experimental values~\cite{kittel}.}
\label{fig:B}
\end{figure}

\begin{figure*}
\includegraphics[width=17.8 cm]{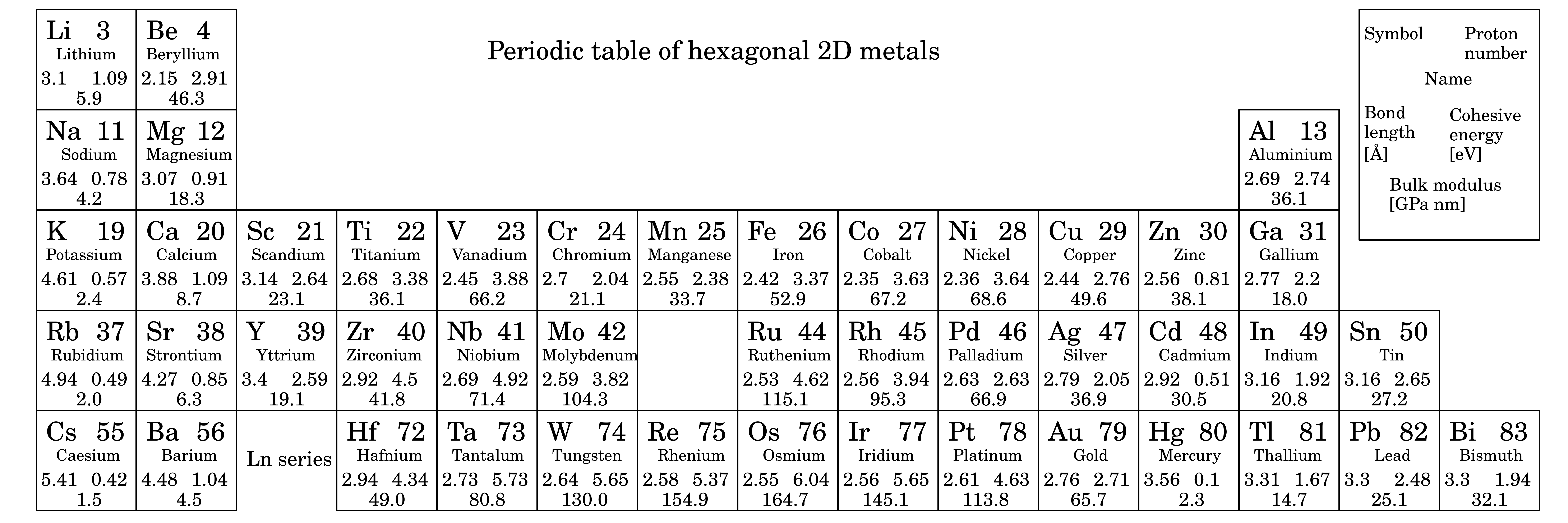}
\caption{Summary of calculated results for hexagonal 2D structures.}
\label{fig:pt2D}
\end{figure*}

\subsection{Bulk moduli correlate between structures}

In this section, we compare the 2D bulk moduli between different structures and find similar correlations to 3D bulk moduli as in the two previous sections for cohesive energies and bond lengths. The bulk moduli are determined from a fit to isotropically deformed structures. Like cohesive energies and bond lengths, the bulk moduli  $B$ correlate between different 2D structures and fittings give \mbox{$B_{sq} = 0.860 \times B_{hex}$}, \mbox{$B_{hc} = 0.576 \times B_{hex}$}, and \mbox{$B_{hc} = 0.666 \times B_{sq}$}. The 2D bulk moduli display roughly similar behavior as a function of the proton number as the cohesion energies (\mbox{Fig. \ref{fig:B}}). The bulk moduli are the largest near the middle of the d-series and increases from 3d-metals to 5d-metals. The bulk moduli for the 2D structures correlate linearly with the experimental 3D bulk moduli. Cr is an exception for this correlation, with experimental 3D bulk modulus of \mbox{$190$ GPa}.

Since the cohesions are the strongest and bond lengths the shortest near the middle of the d-series, it is reasonable that also the 2D bulk moduli are the highest near the middle of d-series. Shorter bonds are harder to contract and elongate, which leads to higher bulk modulus. Note that while the correlations between calculated and experimental cohesive energies and bond lengths have the same units, the bulk moduli have different units in 2D. The 3D bulk moduli are in units GPa, while the 2D bulk moduli are in units GPa nm. The results for hexagonal 2D structures are summarized in \mbox{Fig. \ref{fig:pt2D}}.

\subsection{Elastic constants}

\begin{figure}
\includegraphics[width=8.6 cm]{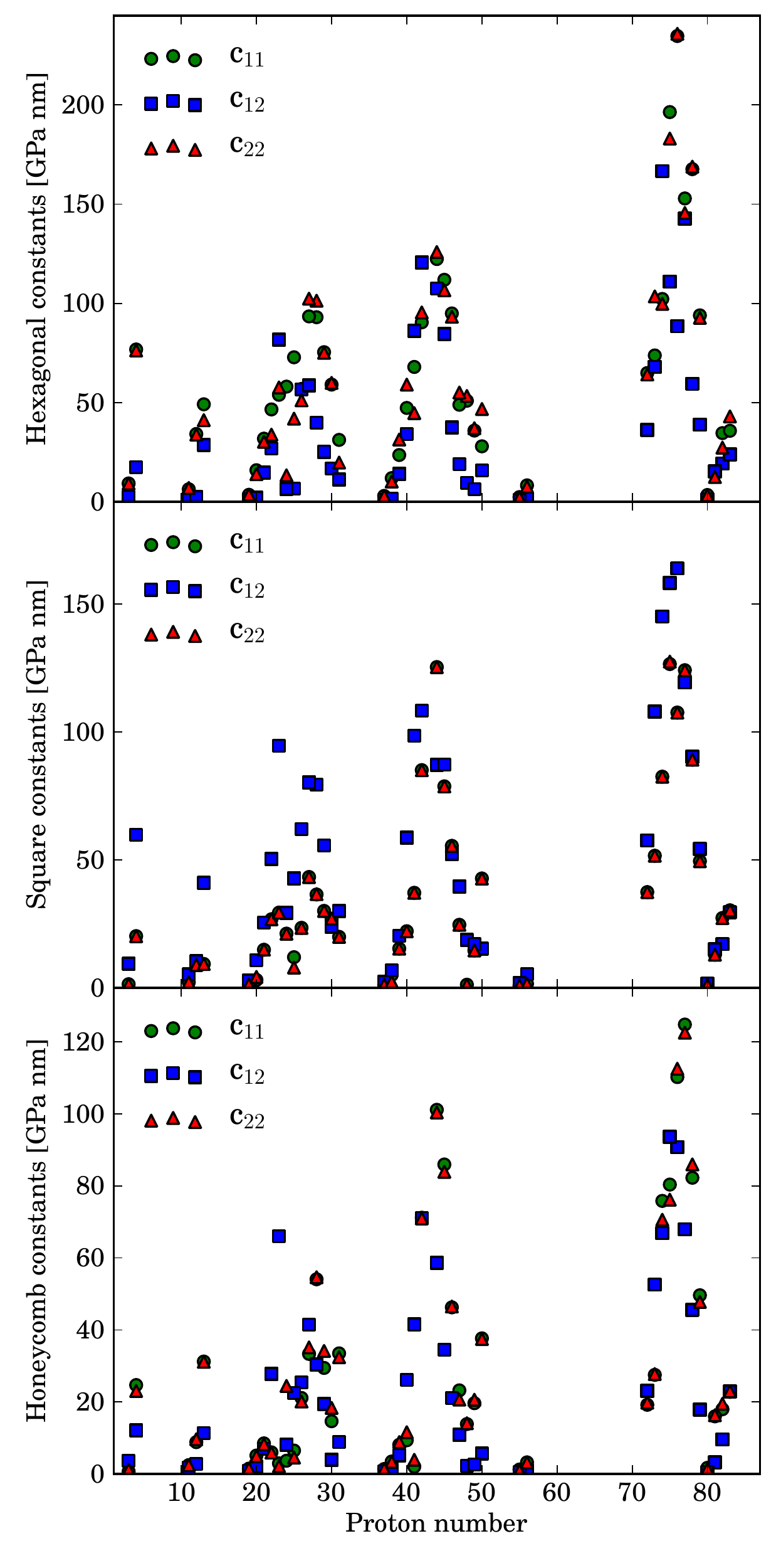}
\caption{Elastic constants for hexagonal, square and honeycomb 2D structures.}
\label{fig:c}
\end{figure}

\begin{figure}
\includegraphics[width=8.6 cm]{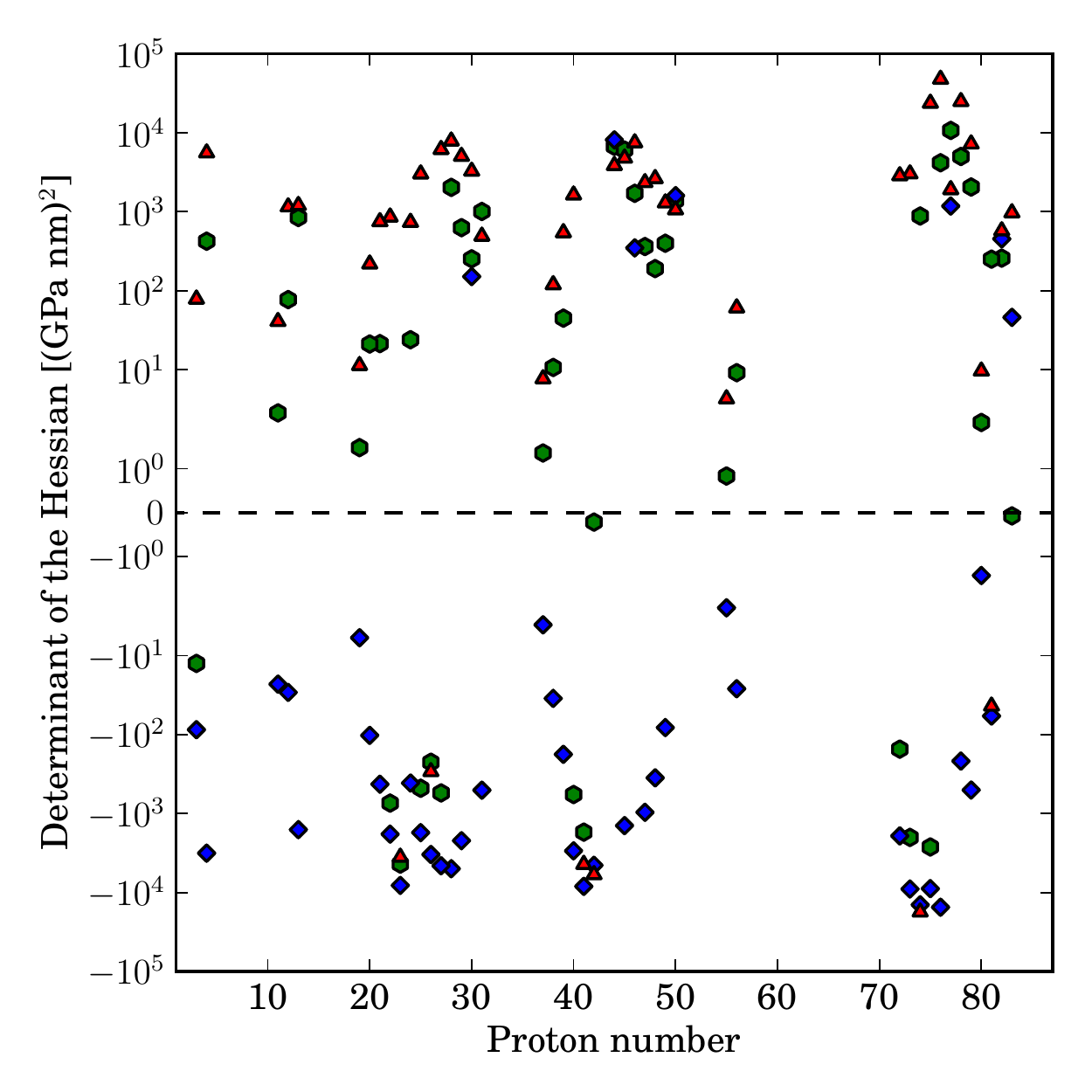}
\caption{Determinant of the Hessian matrix of the energy density $U$. Triangles, squares and hexagons correspond to hexagonal, square, and honeycomb structures, respectively. Logarithmic scale is used for clarity.}
\label{fig:H}
\end{figure}

So far all the considered structures have had ideal geometries, with bond angles $60^\circ$ (hex), $90^\circ$ (sq), and $120^\circ$ (hc), but now we consider also anisotropic strains. We determine the 2D elastic constants $c_{11}$, $c_{12}$, and $c_{22}$ defined by the equation

\begin{equation}
U(\epsilon_x,\epsilon_y) = \frac{1}{2}c_{11}\epsilon_x^2+c_{12}\epsilon_x\epsilon_y+\frac{1}{2}c_{12}\epsilon_y^2,
\end{equation}
where $U$ is the energy density, $\epsilon_x$ the strain in $x$-direction, and $\epsilon_y$ is the strain in $y$-direction around a critical point of the potential energy surface. The elastic constants are obtained by calculating multiple strained structures surrounding the ideal geometry and fitting a second order polynomial to the potential energy surface. The fit has the form

\begin{equation}
U = \frac{c_{11}}{2}x'^2+c_{12}x'y'+\frac{c_{22}}{2}y'^2+U_0,
\end{equation}
where $x'=x-x_0$, $y'=y-y_0$, $x_0$, $y_0$, and $U_0$ are fit parameters, and $x$ and $y$ are strains in $x$- and $y$-directions with respect to the ideal structure (\mbox{Fig. \ref{fig:lattice_sketch}}). The fit parameters $x_0$ and $y_0$ are included to allow for deviations from the ideal geometry in order to center on a critical point of the potential energy surface. Therefore, non-zero values for parameters $x_0$ and $y_0$ indicate that the critical point of the potential energy surface is located on a geometry with non-ideal bonding angles. The energy density $U$ used in the fit is obtained by dividing the energy of the strained structure by the area of the computational cell of the ideal equilibrium structure. The values of the resulting elastic constants are given in the Appendix and shown in \mbox{Fig. \ref{fig:c}}. The general behavior of the elastic constants follows the behavior of the bulk moduli. The values of elastic constants are the highest near the middle of d-series. While the elastic constants  $c_{ij}$ are hard to determine experimentally, they can be used to calculate the Young moduli

\begin{equation}
Y_i = \frac{c_{ii}^2-c_{ij}^2}{c_{ii}},\ \{i,j\} = \{1,2\},\{2,1\}
\end{equation}
and Poisson ratios
\begin{equation}
\nu_i=\frac{c_{ij}}{c_{ii}},\ \{i,j\} = \{1,2\},\{2,1\}
\end{equation}
for the structures, which may be more accessible to experiments.

For the square structures the values of $c_{12}$ are large compared to $c_{11}$ and $c_{22}$. This indicates structural instabilities against in-plane strains. To quantify this observation and to estimate the stability of a given geometry with respect to in-plane deformations, we calculate the determinant of the Hessian matrix of U,

\begin{equation}
|\mathcal{H}| = \begin{vmatrix} \frac{\partial^2 U}{\partial^2 x} & \frac{\partial^2 U}{\partial x \partial y} \\ \frac{\partial^2 U}{\partial y \partial x} & \frac{\partial^2 U}{\partial^2 y} \end{vmatrix} = c_{11}c_{22}-c_{12}^2.
\end{equation}
If $|\mathcal{H}|$ is negative, the critical point close to the ideal structure is a saddle point and the geometry is unstable. The values for $|\mathcal{H}|$ for studied metals and structures are shown in \mbox{Fig. \ref{fig:H}}. For most metals in the hexagonal structure, $|\mathcal{H}|$ is positive, and structures thus stable. Many metals are stable also in the honeycomb structure. However, despite few exceptions, most metals in the square structures are unstable.

\subsection{Bending modulus}

\begin{figure}
\includegraphics[width=8.6 cm]{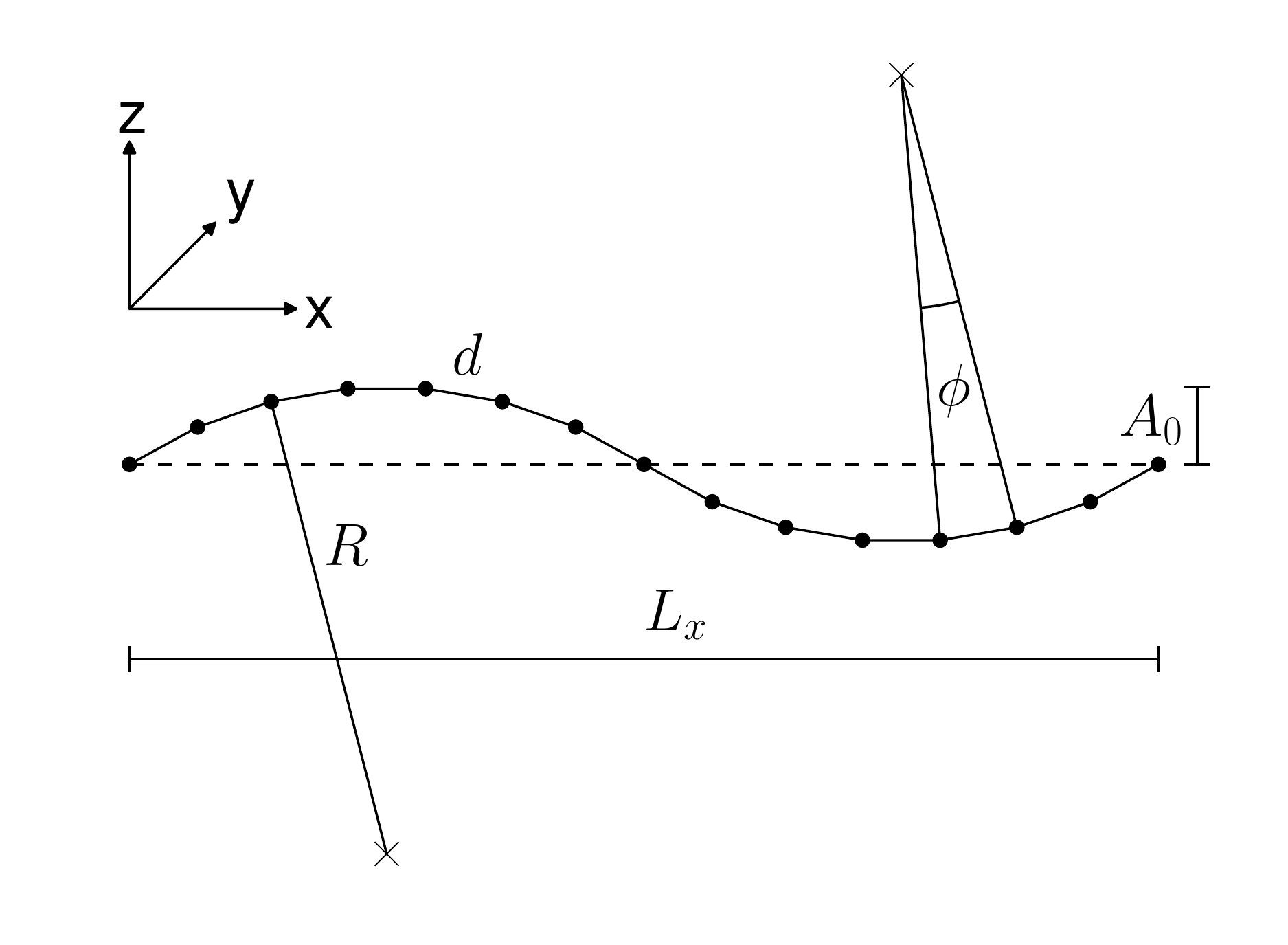}
\caption{Schematic representation of the structures used in the calculation of bending moduli.}
\label{fig:bending}
\end{figure}

\begin{figure}
\includegraphics[width=8.6 cm]{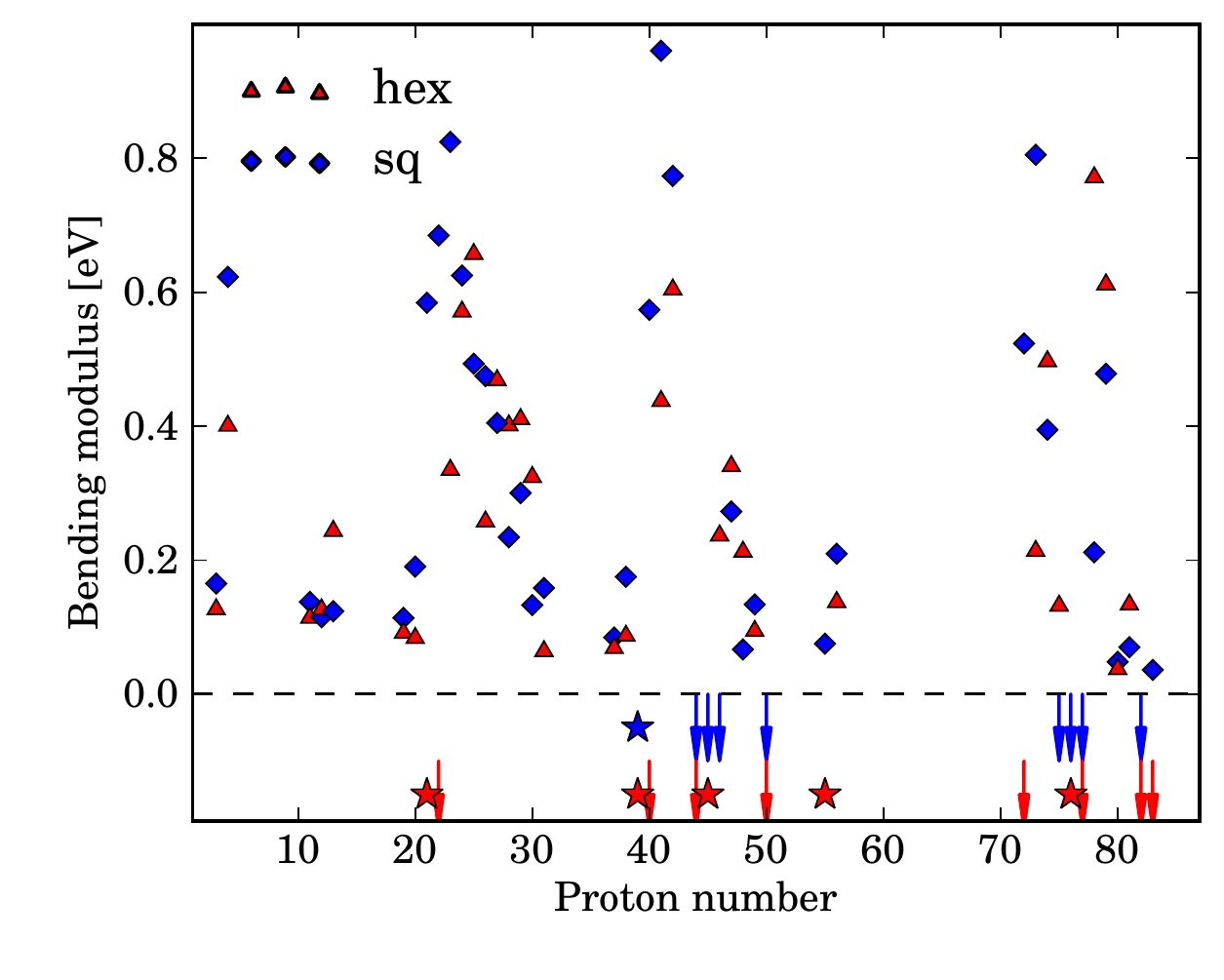}
\caption{Calculated bending moduli for the hexagonal (hex) and square (sq) structures. Arrows indicate negative bending moduli and stars undetermined values for hexagonal (red) and square (blue) 2D structures.}
\label{fig:kappa}
\end{figure}

Thus far, all deformations have been restricted to the atomic plane, but now the atoms are displaced also out of plane. We introduce deformations to calculate the bending moduli for the hexagonal and square lattices. To regain periodicity, we construct bent structures consisting of two connected cylindrical sections (\mbox{Fig. \ref{fig:bending}}). All the atoms are located on cylindrical surfaces, yielding constant radii of curvature throughout the entire structure, with the bond lengths fixed to the flat equilibrium values. The method ignores possible in-plane strains, caused by expansion or contraction due to bending. However, for most metals and small curvatures this contribution is small, as evidenced by fair agreement with energy density fits that are based on pure bending alone. The construction of a bent structure for a given radius of curvature $R$ and bond length $d$ is as follows: the angle of rotation between neighboring atoms is

\begin{equation}
\phi = 2\mathrm{\ arcsin}\left(\frac{d}{2R}\right)
\end{equation} 
and the length of the required computational cell is
\begin{equation}
L_x = 4R\mathrm{\ sin}\left(\frac{N_x\phi}{4}\right),
\end{equation} 
where $N_x$ is the number of atoms in the $x$-direction required to make the full bend. Here, we used 14 atoms for the bent square lattices. The bending amplitude is given by

\begin{equation}
A_0 = R\left[1-\mathrm{cos}(N_x\phi/4)\right],
\end{equation}
and the centers of the cylinders by the coordinates $(L_x/4,0,R-A_0)$ and $(3L_x/4,0,A_0-R)$. The structures are then constructed by rotating atoms around the centers of the cylinders by the angle $\phi$, until the geometry shown in \mbox{Fig. \ref{fig:bending}} is obtained. To get the bent hexagonal structures, a second row of atoms is added, with atoms rotated by $\phi/2$ at a $y$-distance 

\begin{equation}
y_{\mathrm{hex}}/2 = \sqrt{3d^2/4-R^2\left[1-\mathrm{cos}(\phi/2)\right]^2}.
\end{equation}
The directions are indicated by the axes in \mbox{Fig. \ref{fig:bending}}. In the hexagonal structures, one bond at the connection point between the cylinders is slightly strained, but the effect is small enough to be neglected. The bending modulus $\kappa$ is then obtained by fitting the energy density as

\begin{equation}
U(R) = \frac{1}{2}\frac{\kappa}{R^2}+U_0.
\end{equation}
The energy density is \mbox{$U(R)=E(R)/A(R)$}, where $E(R)$ is the energy of the bent structure with radius of curvature $R$, and $A(R)$ is the area of the ideal cylindrical surface, given by \mbox{$A_{\mathrm{sq}}(R)=N_x\phi R d$} for the square and \mbox{$A_{\mathrm{hex}}(R)=N_x\phi R y_{\mathrm{hex}}$} for the hexagonal structures. The bending moduli thus obtained are shown in \mbox{Fig. \ref{fig:kappa}}. Most metals have positive bending moduli, indicating stability against bending. The bending moduli in 3d-series have behavior qualitatively similar to the bulk moduli: the bending moduli are the highest near the middle of the series, corresponding to short bond lengths. For the 4d- and 5d-series, however, the behavior is different. This time negative values for bending moduli appear near the middle of the series. Most important, the bending moduli are surprisingly large, some of them even around\mbox{\ $\sim 1$ eV}, and thus well comparable to bending modulus in graphene \mbox{($\kappa = 1$ eV)} and other covalently bound 2D materials~\cite{koskinen10,koskinen14}.

\subsection{Electron density changes mostly in plane}

\begin{figure}
\includegraphics[width=8.6 cm]{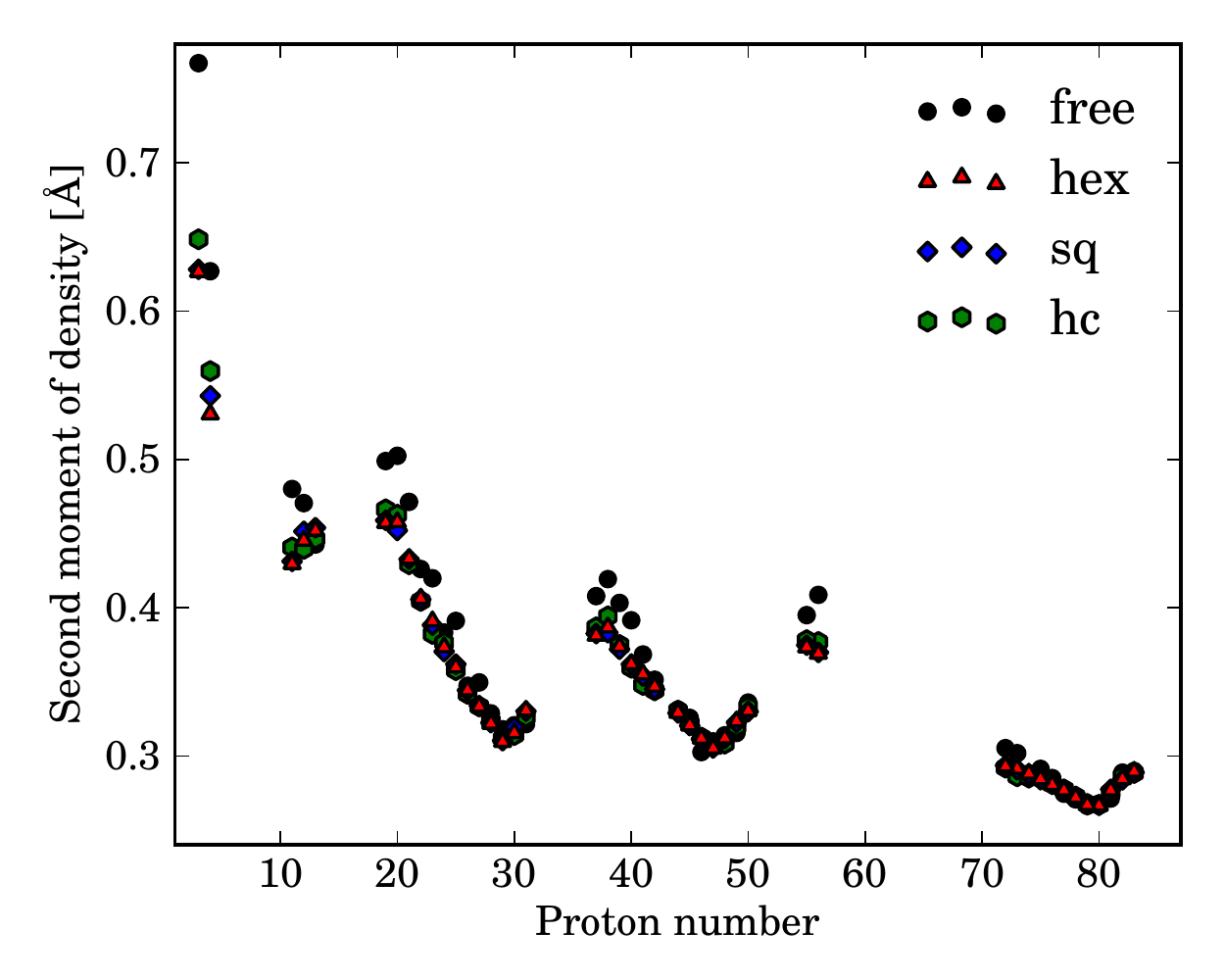}
\caption{The second moments of electron density in the direction perpendicular to the atomic plane of the 2D structures.}
\label{fig:z}
\end{figure}

Last we consider how the 2D structure affects the electron density. In this section, we again consider ideal equilibrium structures. We begin by investigating the electron density integrated perpendicular across the atomic plane. Surprisingly, the electron density perpendicular to atomic plane remains nearly unaffected by the choice of 2D geometry. To quantify this observation, we computed the second moment of electron density $n(\vec{r})$ for free atoms and the different 2D structures

\begin{equation}
\langle z^2 \rangle^\frac{1}{2} = \Big(\int\mathrm{d}\mathbf{r} z^2 n(\mathbf{r})\bigg/\int\mathrm{d}\mathbf{r} n(\mathbf{r})\Big)^\frac{1}{2},
\end{equation}
where $z$-direction is perpendicular to the atomic plane. These results are summarized in \mbox{Fig. \ref{fig:z}}. With the exception of the lightest metals Li and Be, the second moment is almost independent of the 2D structure. However, mostly the second moment is reduced from the value of the free atoms, indicating the relocation of charge from the more diffuse free atom states to the in-plane bonding regions between the atoms. For the simple metals the second moment reduces the most while for the transition metals it can even increase, as compared to free atoms. This is probably due to spin structures, which are more complicated in free transition metal atoms.

\begin{figure}
\includegraphics[width=8.6 cm]{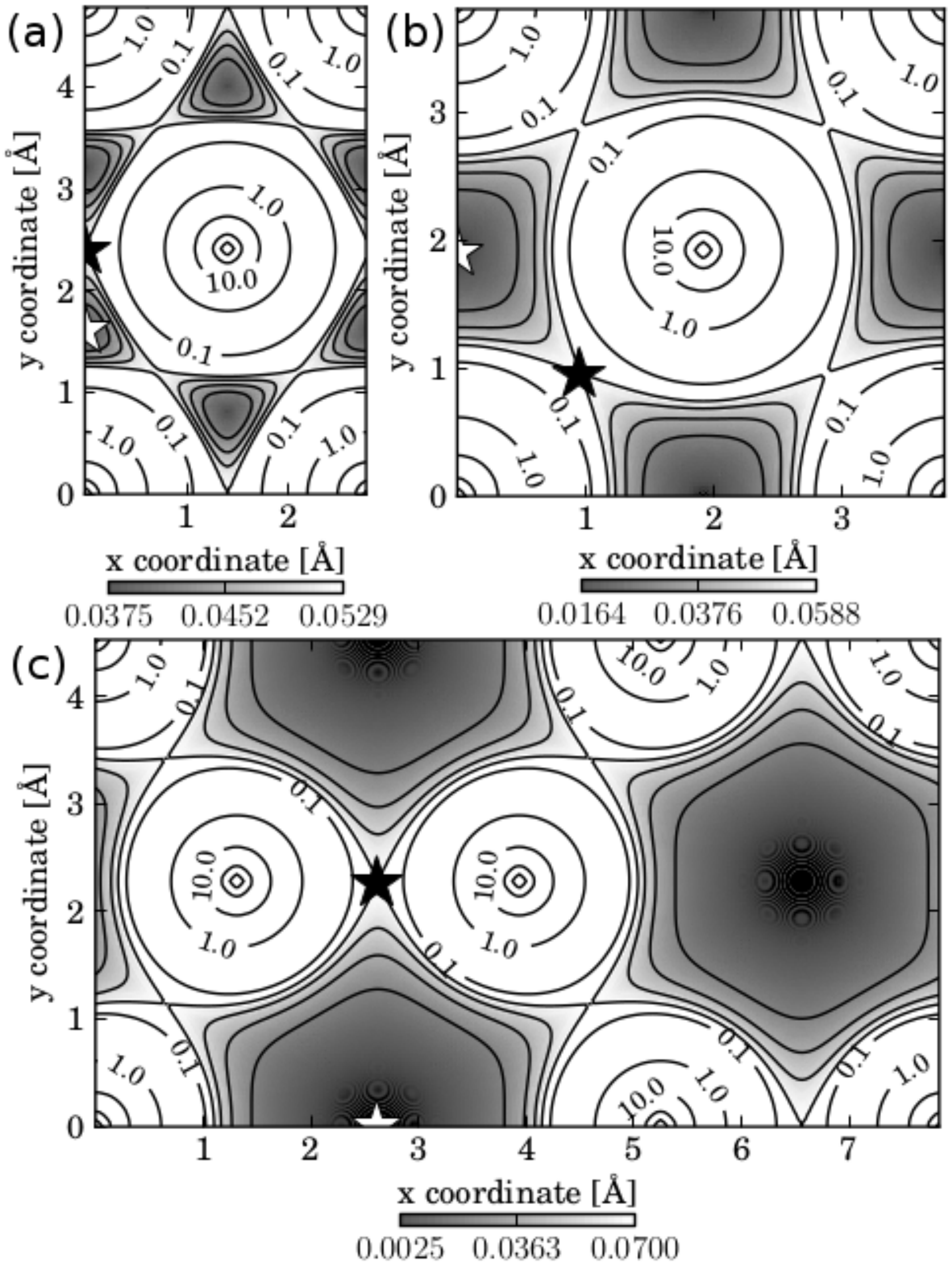}
\caption{The electron density of Au in the atomic plane, for hexagonal (a), square (b), and honeycomb (c) latices, in atomic units. Black star shows the position of saddle point and white star shows the location of a minimum point.}
\label{fig:n2D}
\end{figure}

Let us proceed to investigate the electron density in the atomic planes. \mbox{Fig. \ref{fig:n2D}} shows the electron density of Au for the three different structures. In all three cases the electron density has a minimum at the corner of the Wigner-Seitz cell. In addition to the large maxima at the positions of the nuclei, there are also saddle points between the atoms in all structures. The arrangement of these critical points are rather similar in all studied metals. Further, these critical points of the electron density are located in the atomic plane.

To quantify the changes in the electron density between different metals and structures, we calculate the ratios between the densities in the local minima at the corners of the Wigner-Seitz cells and at the saddle points half-way between the atoms. This measure has been used to characterize the metallicity of a bond: the higher the ratio, the flatter the electron density in the bonding region and the more metallic the bond~\cite{mori02}.

\begin{figure}
\includegraphics[width=8.6 cm]{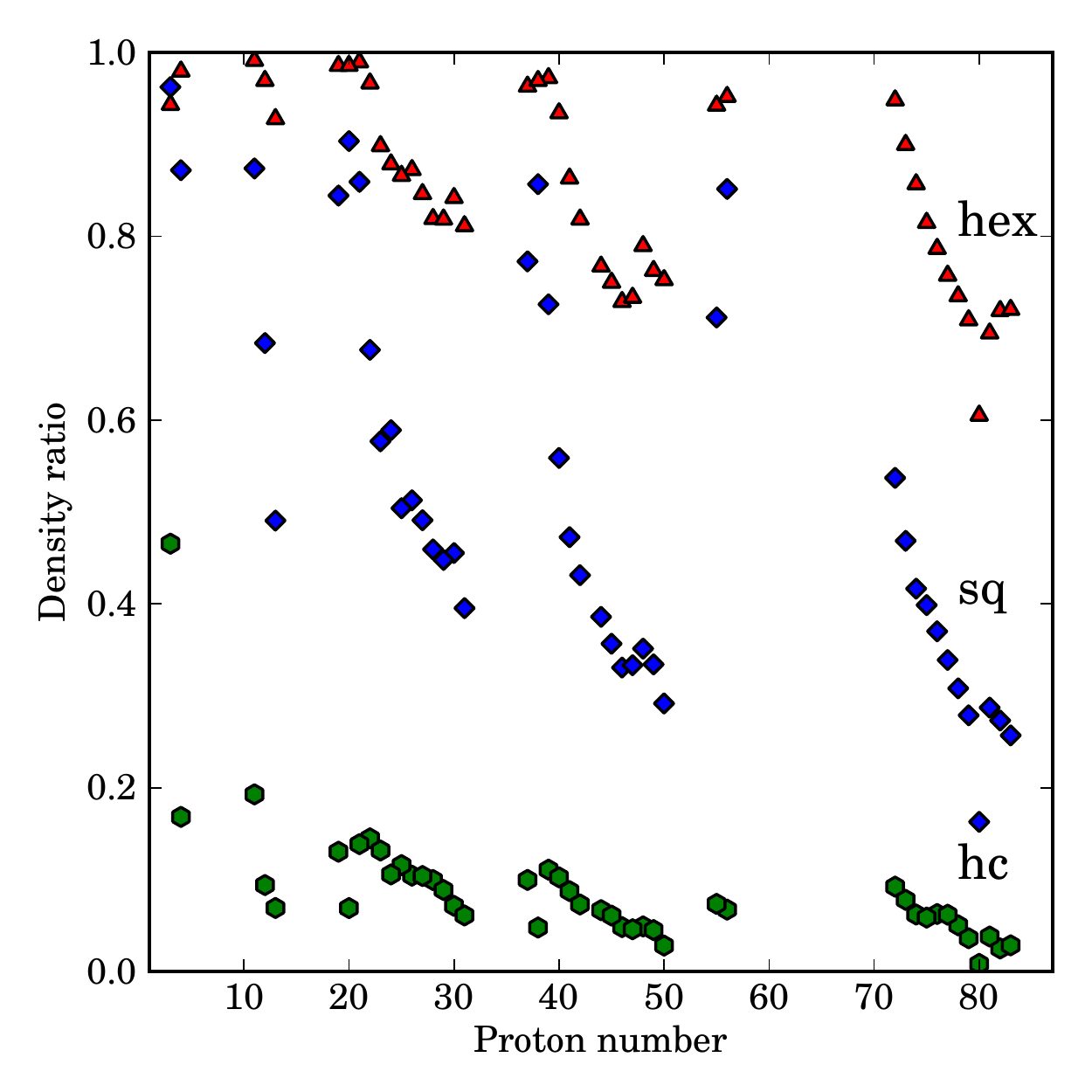}
\caption{The ratio between minimum electron density and the saddle point density at the atomic plane, for hexagonal (hex), square (sq), and honeycomb (hc) lattices.}
\label{fig:f}
\end{figure}

The resulting density ratios are larger for the simple metals than for the transition metals (\mbox{Fig. \ref{fig:f}}). This is probably due to the more localized nature of the d-electrons: While the s-and p-electrons delocalize and produce smoother electron density, the d-electrons remain partly localized and produce a larger difference in the electron density in the bonding region between the atoms. An even clearer trend emerges between the structures. In general, the ratio decreases when the number of nearest neighbors decreases. The ratio is largest for hexagonal lattice, intermediate for square lattice, and smallest for the honeycomb lattice. The behavior in density ratio is related to the geometries of the Wigner-Seitz cells of the different structures. While the hexagonal and square lattices are rather densely packed, the honeycomb lattice is sparser and has large distance between nuclei and the corners of the Wigner-Seitz cell. Therefore, the electron density ratio for honeycomb is low. Large density ratios can be associated with metallic bonding and small density ratio values with covalent bonding. We note that the density ratios of square structures are intermediate between metallic and covalent bonding, which could therefore partly explain the instability of the square structures. While changing the geometry from hexagonal to square lattice decreases the density ratio and weakens the metallic bonds, it still fails to fully lead to covalent bonding as in the honeycomb structure.

\begin{figure}
\includegraphics[width=8.6 cm]{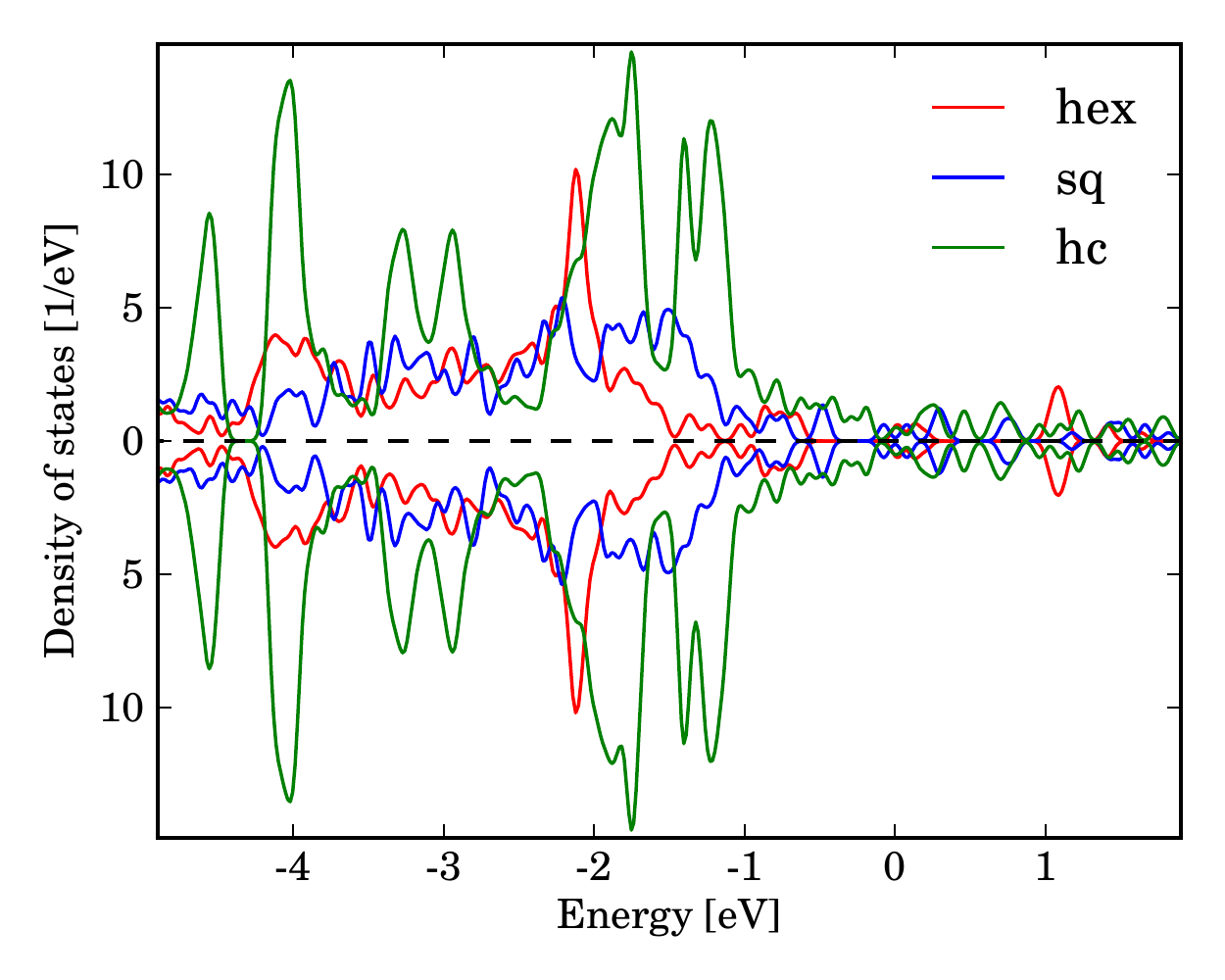}
\caption{Density of states for different 2D Au structures.}
\label{fig:dos}
\end{figure}

\begin{figure}
\includegraphics[width=8.6 cm]{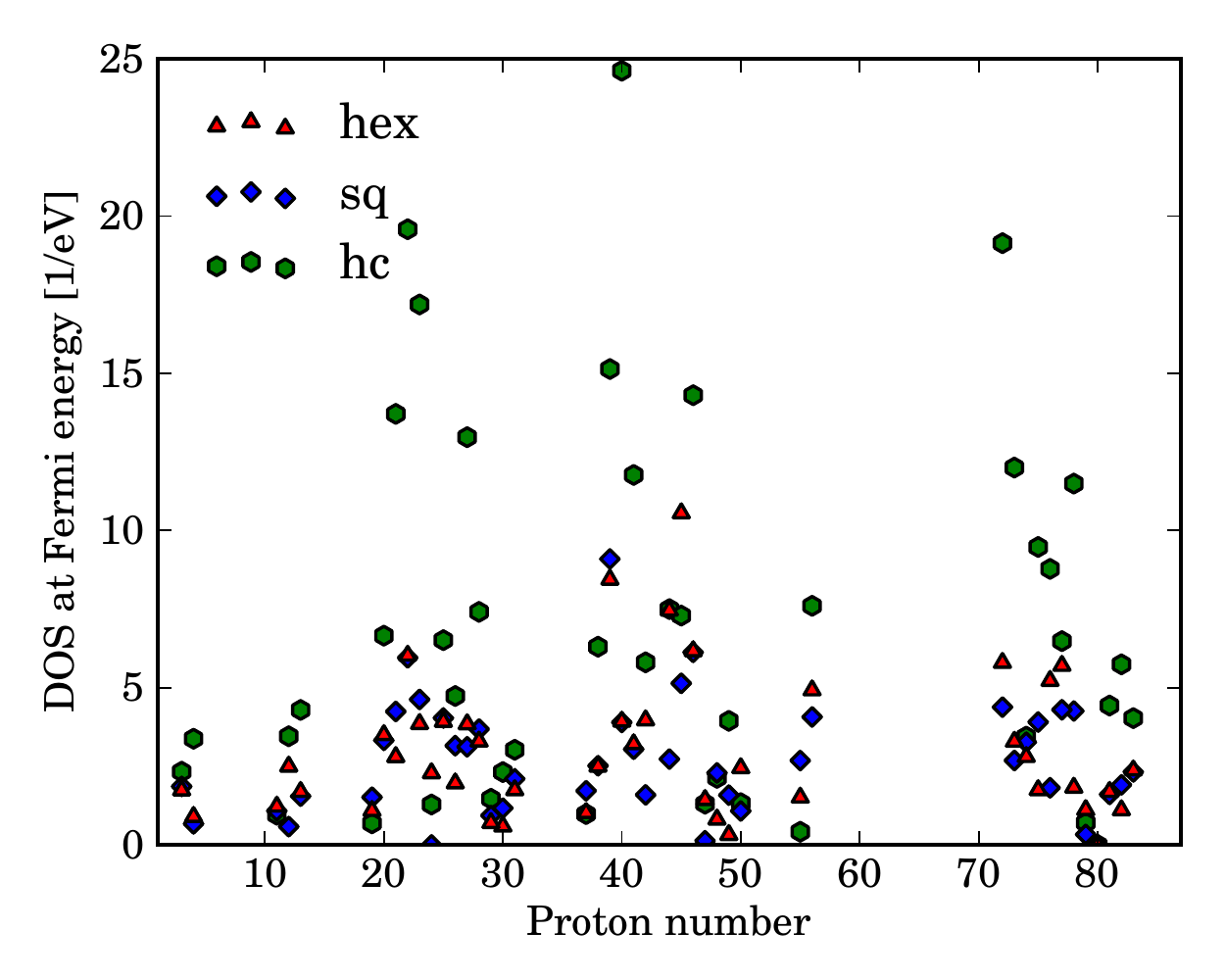}
\caption{Density of states at Fermi energy for all studied metals and 2D structures.}
\label{fig:dos_ef}
\end{figure}

Finally, we consider the density of states (DOS) of the 2D structures, using Au as an example (\mbox{Fig. \ref{fig:dos}}). We use Gaussian smearing of 0.1 eV and set Fermi energy to zero for each geometry. Spin-up densities are shown above zero and spin-down densities are shown below zero. Some qualitative differences are observed between the different geometries. The hexagonal structure has a large peak near \mbox{$-2$ eV}, a feature not present in the square structure. In the square geometry, the DOS is flatter and does not have clear peaks. The honeycomb structure has DOS more featured and peaked than those of the hexagonal and square geometries. In order to compare the electronic structures between different metals and structures, we calculated the DOS at Fermi energy (\mbox{Fig. \ref{fig:dos_ef}}). Most metals have nonzero DOS at Fermi energy, as expected. No general trends, however, appear in these DOS values.

\section{Summary and Conclusions}

We studied semi-infinite free-standing mono-atomic metal layers with density-functional calculations. We found that the cohesive energies, bond lengths, and bulk moduli in the hexagonal, square, and honeycomb 2D geometries correlate linearly with the corresponding 3D bulk values. Therefore, the same properties correlate linearly also between the 2D structures. The hexagonal structures, in general, have the highest cohesion, longest bond lengths, and highest bulk moduli. Moreover, the cohesive energies are the highest, bond lengths the shortest, and 2D bulk moduli the highest near the middle of the d-series. The trend in cohesion was associated with the d-band filling. The half-filled d-band in the middle of the d-series corresponds to filled bonding orbitals and empty anti-bonding orbitals. Simple metals lack the d-electron contribution to cohesion and therefore have smaller cohesive energies. Further, stronger bonds lead to shorter bond lengths and higher bulk moduli. Since cohesive energies correlate between 2D and 3D bulk structures, choosing a suitable metal for a stable 2D structure faces a trade-off between sufficiently high 2D cohesion and sufficiently low 3D bulk cohesion. Too low 2D cohesion renders the 2D structure unstable, while too high 3D bulk cohesion leads to growth of a 3D structure.

The calculated in-plane elastic constants indicate that most metals are stable in the hexagonal and honeycomb geometries, but unstable in the square lattice geometry. Analysis of the bending moduli showed that many metals are stable against bending, even in the square geometry. Since all the structures considered in this Article are semi-infinite, determining whether a certain geometry is stable when placed inside a 2D support structure, such as a graphene nanopore, is not directly accessible using only the values and perspectives presented here. For example, bonding to a support structure might change the stability of a given geometry due to charge transfer between support and metal. Still, the effects of strain to the total energy of a combined 2D metal in a support structure system can be estimated using the calculated elastic moduli. For instance, since graphene is extremely stiff against compression in the atomic plane, most of the deformations due to lattice mismatch will occur in the 2D metal, not in graphene. The change in energy due to this strain can be approximated using our values for semi-infinite 2D metals.

Electron density analysis showed that changing the 2D structure leaves the perpendicular density profile nearly unchanged. This was quantified by calculating the second moment of electron density in the direction perpendicular to the atomic plane. While the second moment of the electron density changes during the formation of a 2D structure from free atoms, it remains nearly constant when the 2D geometry is changed between hexagonal, square, and honeycomb lattices. Yet in the atomic plane the electron density does change. This change was quantified by studying the critical points of the density in the plane of the atoms. The electron density is smoother for the structures with more nearest neighbors, as indicated by the larger ratios between electron densities at the minima far from the atomic nuclei and the saddle points half-way between the atoms. Decrease in this density ratio is related to the increase of covalency of the bonds. All these energetic, geometric, elastic, and electronic trends, tabulated also in the Appendix for completeness and future reference, make up an atlas that provides essential guidance in the research involving supported and free-standing 2D metal nanostructures.

\section*{Acknowledgments}

We acknowledge the Academy of Finland for funding (project 297115).

\section*{Appendix}

\begin{center}
\begin{table*}
\caption{Cohesive energies E (eV), bonding lengths d (\AA), bulk moduli B (GPa nm), and elastic constants \mbox{(GPa nm)} for hexagonal, square, and honeycomb structures.}
\begin{tabular}{c|cccccc|cccccc|cccccc}

 & \multicolumn{6}{c|}{Hexagonal} & \multicolumn{6}{c|}{Square} & \multicolumn{6}{c}{Honeycomb} \\
 & E & d & B & c11 & c12 & c22 & E & d & B & c11 & c12 & c22 & E & d & B & c11 & c12 & c22 \\
\hline
Ag & 2.05 & 2.79 & 36.9 & 48.9 & 18.9 & 55.1 & 1.83 & 2.71 & 30.0 & 24.6 & 39.6 & 24.6 & 1.5 & 2.67 & 16.9 & 23.2 & 10.9 & 20.7 \\
Al & 2.74 & 2.69 & 36.1 & 49.1 & 28.7 & 41.2 & 2.46 & 2.65 & 25.8 & 9.3 & 41.0 & 9.3 & 2.25 & 2.59 & 21.1 & 31.2 & 11.3 & 31.2 \\
Au & 2.71 & 2.76 & 65.7 & 93.9 & 38.9 & 92.8 & 2.41 & 2.69 & 51.4 & 49.5 & 54.4 & 49.5 & 2.06 & 2.61 & 34.4 & 49.6 & 17.9 & 47.8 \\
Ba & 1.04 & 4.48 & 4.5 & 8.3 & 1.6 & 7.7 & 0.9 & 4.24 & 3.5 & 1.5 & 5.4 & 1.6 & 0.52 & 4.47 & 2.1 & 3.2 & 1.0 & 3.2 \\
Be & 2.91 & 2.15 & 46.3 & 76.8 & 17.4 & 76.3 & 2.57 & 2.02 & 40.3 & 20.2 & 59.8 & 20.2 & 1.61 & 2.13 & 18.1 & 24.7 & 12.1 & 23.0 \\
Bi & 1.94 & 3.3 & 32.1 & 35.8 & 23.9 & 43.1 & 1.97 & 3.13 & 29.5 & 30.3 & 29.5 & 30.3 & 2.0 & 3.05 & 23.4 & 22.8 & 22.9 & 22.9 \\
Ca & 1.09 & 3.88 & 8.7 & 15.9 & 2.0 & 13.9 & 0.91 & 3.65 & 6.6 & 3.1 & 10.7 & 4.3 & 0.57 & 3.88 & 3.5 & 5.1 & 2.0 & 4.9 \\
Cd & 0.51 & 2.92 & 30.5 & 50.9 & 9.5 & 53.4 & 0.34 & 2.92 & 9.4 & 1.2 & 18.8 & 1.2 & 0.24 & 2.92 & 7.4 & 13.8 & 2.2 & 14.1 \\
Co & 3.63 & 2.35 & 67.2 & 93.4 & 58.6 & 102.4 & 3.35 & 2.25 & 59.1 & 43.3 & 80.3 & 43.3 & 2.79 & 2.18 & 38.1 & 33.3 & 41.5 & 35.2 \\
Cr & 2.04 & 2.7 & 21.1 & 58.0 & 6.4 & 13.4 & 2.1 & 2.41 & 24.7 & 21.2 & 29.3 & 21.2 & 1.1 & 2.68 & 11.5 & 3.7 & 8.1 & 24.4 \\
Cs & 0.42 & 5.41 & 1.5 & 2.4 & 0.8 & 2.1 & 0.37 & 5.11 & 1.4 & 0.9 & 1.8 & 0.9 & 0.29 & 5.05 & 0.8 & 1.2 & 0.5 & 0.9 \\
Cu & 2.76 & 2.44 & 49.6 & 75.4 & 25.2 & 75.1 & 2.45 & 2.37 & 40.1 & 30.1 & 55.7 & 30.1 & 2.06 & 2.3 & 25.1 & 29.4 & 19.4 & 34.2 \\
Fe & 3.37 & 2.42 & 52.9 & 56.9 & 56.6 & 51.1 & 3.04 & 2.32 & 42.3 & 23.5 & 62.0 & 23.5 & 2.36 & 2.27 & 23.3 & 21.1 & 25.5 & 20.1 \\
Ga & 2.2 & 2.77 & 18.0 & 31.2 & 11.2 & 19.8 & 2.13 & 2.64 & 24.0 & 19.9 & 30.0 & 19.9 & 1.99 & 2.52 & 20.9 & 33.5 & 8.9 & 32.4 \\
Hf & 4.34 & 2.94 & 49.0 & 64.9 & 36.2 & 64.2 & 4.0 & 2.79 & 46.2 & 37.4 & 57.6 & 37.4 & 2.74 & 2.82 & 20.8 & 19.2 & 23.1 & 19.8 \\
Hg & 0.1 & 3.56 & 2.3 & 3.4 & 1.2 & 3.3 & 0.07 & 3.54 & 1.4 & 1.1 & 1.7 & 1.2 & 0.06 & 3.47 & 1.0 & 1.7 & 0.5 & 1.4 \\
In & 1.92 & 3.16 & 20.8 & 35.9 & 6.3 & 37.1 & 1.85 & 3.01 & 15.2 & 14.6 & 17.2 & 14.6 & 1.71 & 2.88 & 11.0 & 19.6 & 2.6 & 20.6 \\
Ir & 5.65 & 2.56 & 145.1 & 152.9 & 142.7 & 145.6 & 5.25 & 2.45 & 123.4 & 124.2 & 119.3 & 124.2 & 5.04 & 2.35 & 96.3 & 124.8 & 67.9 & 122.6 \\
K & 0.57 & 4.61 & 2.4 & 3.5 & 1.1 & 3.6 & 0.53 & 4.34 & 2.0 & 1.4 & 2.8 & 1.4 & 0.41 & 4.28 & 1.2 & 1.5 & 0.9 & 1.5 \\
Li & 1.09 & 3.1 & 5.9 & 9.2 & 2.6 & 9.2 & 1.02 & 2.9 & 5.3 & 1.4 & 9.4 & 1.4 & 0.76 & 2.77 & 2.3 & 0.7 & 3.7 & 1.2 \\
Mg & 0.91 & 3.07 & 18.3 & 34.2 & 2.3 & 33.9 & 0.69 & 2.96 & 9.3 & 8.9 & 10.4 & 8.9 & 0.46 & 3.05 & 5.9 & 8.8 & 2.8 & 9.6 \\
Mn & 2.38 & 2.55 & 33.7 & 72.8 & 6.7 & 41.9 & 2.23 & 2.28 & 25.4 & 12.0 & 42.8 & 8.0 & 1.58 & 2.44 & 12.0 & 6.5 & 22.5 & 4.6 \\
Mo & 3.82 & 2.59 & 104.3 & 90.5 & 120.6 & 95.5 & 3.7 & 2.45 & 97.1 & 85.1 & 108.4 & 85.1 & 3.19 & 2.33 & 71.7 & 71.2 & 71.1 & 71.0 \\
Na & 0.78 & 3.64 & 4.2 & 6.3 & 1.8 & 7.0 & 0.73 & 3.43 & 3.6 & 2.2 & 5.3 & 2.2 & 0.58 & 3.37 & 2.2 & 2.4 & 1.8 & 2.4 \\
Nb & 4.92 & 2.69 & 71.4 & 68.0 & 86.2 & 44.7 & 4.6 & 2.56 & 69.4 & 37.1 & 98.5 & 37.1 & 3.55 & 2.46 & 24.0 & 2.0 & 41.5 & 3.9 \\
Ni & 3.64 & 2.36 & 68.6 & 93.0 & 39.8 & 101.4 & 3.28 & 2.27 & 55.5 & 36.5 & 79.5 & 36.7 & 2.88 & 2.19 & 41.4 & 54.1 & 30.4 & 54.6 \\
Os & 6.04 & 2.55 & 164.7 & 234.7 & 88.5 & 235.9 & 5.7 & 2.42 & 134.4 & 107.6 & 164.0 & 107.6 & 5.35 & 2.33 & 102.4 & 110.2 & 90.8 & 112.5 \\
Pb & 2.48 & 3.3 & 25.1 & 34.6 & 19.2 & 27.3 & 2.36 & 3.13 & 22.7 & 27.3 & 17.1 & 27.3 & 2.19 & 2.97 & 14.2 & 18.0 & 9.6 & 19.5 \\
Pd & 2.63 & 2.63 & 66.9 & 94.9 & 37.4 & 93.3 & 2.34 & 2.54 & 54.5 & 55.5 & 52.3 & 55.5 & 1.95 & 2.48 & 34.1 & 46.3 & 21.1 & 46.5 \\
Pt & 4.63 & 2.61 & 113.8 & 167.6 & 59.5 & 168.9 & 4.13 & 2.53 & 89.7 & 89.2 & 90.4 & 89.2 & 3.75 & 2.43 & 64.6 & 82.3 & 45.6 & 86.0 \\
Rb & 0.49 & 4.94 & 2.0 & 2.9 & 1.0 & 2.9 & 0.45 & 4.66 & 1.7 & 1.2 & 2.3 & 1.2 & 0.36 & 4.59 & 1.0 & 1.4 & 0.7 & 1.4 \\
Re & 5.37 & 2.58 & 154.9 & 196.4 & 111.0 & 183.1 & 5.02 & 2.45 & 132.1 & 126.6 & 158.3 & 127.5 & 4.43 & 2.35 & 86.4 & 80.4 & 93.6 & 76.2 \\
Rh & 3.94 & 2.56 & 95.3 & 111.9 & 84.6 & 106.6 & 3.78 & 2.44 & 82.0 & 78.8 & 87.3 & 78.8 & 3.44 & 2.35 & 58.9 & 85.9 & 34.5 & 83.9 \\
Ru & 4.62 & 2.53 & 115.1 & 122.3 & 107.5 & 125.9 & 4.66 & 2.38 & 107.4 & 125.4 & 87.2 & 125.4 & 4.28 & 2.29 & 80.0 & 101.1 & 58.7 & 100.4 \\
Sc & 2.64 & 3.14 & 23.1 & 31.8 & 14.6 & 30.3 & 2.41 & 2.93 & 23.8 & 14.9 & 25.4 & 15.0 & 1.44 & 2.99 & 7.9 & 8.5 & 7.0 & 8.3 \\
Sn & 2.65 & 3.16 & 27.2 & 27.9 & 15.8 & 46.8 & 2.63 & 2.95 & 29.8 & 42.8 & 15.3 & 42.8 & 2.6 & 2.77 & 22.8 & 37.6 & 5.7 & 37.5 \\
Sr & 0.85 & 4.27 & 6.3 & 11.9 & 1.5 & 10.2 & 0.68 & 4.02 & 5.1 & 5.0 & 6.8 & 2.3 & 0.4 & 4.32 & 2.1 & 3.4 & 1.0 & 3.4 \\
Ta & 5.73 & 2.73 & 80.8 & 73.7 & 68.0 & 103.5 & 5.25 & 2.6 & 79.5 & 51.6 & 108.0 & 51.6 & 4.01 & 2.54 & 41.1 & 27.5 & 52.6 & 27.8 \\
Ti & 3.38 & 2.68 & 36.1 & 46.5 & 26.8 & 33.9 & 3.18 & 2.53 & 39.1 & 26.8 & 50.4 & 26.8 & 1.98 & 2.48 & 16.3 & 6.0 & 27.8 & 5.9 \\
Tl & 1.67 & 3.31 & 14.7 & 15.2 & 15.3 & 12.5 & 1.62 & 3.14 & 13.4 & 13.0 & 15.1 & 13.0 & 1.45 & 3.02 & 9.5 & 16.0 & 3.2 & 16.3 \\
V & 3.88 & 2.45 & 66.2 & 53.9 & 81.7 & 57.7 & 3.71 & 2.3 & 62.0 & 29.4 & 94.7 & 29.4 & 2.59 & 2.22 & 32.8 & 2.9 & 66.0 & 2.2 \\
W & 5.65 & 2.64 & 130.0 & 102.2 & 166.7 & 99.8 & 5.22 & 2.5 & 113.3 & 82.5 & 145.2 & 82.5 & 4.48 & 2.39 & 70.3 & 75.8 & 66.9 & 70.6 \\
Y & 2.59 & 3.4 & 19.1 & 23.6 & 14.0 & 31.4 & 2.41 & 3.22 & 17.2 & 15.3 & 20.3 & 15.3 & 1.54 & 3.29 & 6.9 & 8.1 & 5.2 & 8.9 \\
Zn & 0.81 & 2.56 & 38.1 & 59.1 & 16.8 & 59.9 & 0.56 & 2.49 & 22.3 & 26.6 & 23.9 & 27.1 & 0.37 & 2.53 & 10.4 & 14.6 & 4.0 & 18.4 \\
Zr & 4.5 & 2.92 & 41.8 & 47.3 & 34.1 & 59.1 & 4.17 & 2.79 & 40.0 & 22.1 & 58.7 & 22.1 & 2.96 & 2.73 & 18.3 & 9.3 & 26.1 & 11.6 

\end{tabular}
\end{table*}
\end{center}

\end{document}